\documentclass[11pt,twoside]{article}

\usepackage{asp2006}
\usepackage{epsf}
\usepackage{psfig}
\usepackage{lscape}
\usepackage{graphicx}

\begin{document}
\title{First performances of the GOLF-NG instrumental prototype observing the Sun in Tenerife}   
\author{D. Salabert$^1$, S. Turck-Chi\`eze$^2$, J.~C. Barri\`ere$^2$, 
P.~H. Carton$^2$, P. Daniel-Thomas$^2$, A. Delbart$^2$, R.~A. Garc\'ia$^2$,
R. Granelli$^2$, S.~J. Jim\'enez-Reyes$^1$, C. Lahonde-Hamdoun$^2$, 
D. Loiseau$^2$, S. Mathur$^{2,3}$, F. Nunio$^2$, P.~L. Pall\'e$^1$, Y. Piret$^2$,
J.~M. Robillot$^4$, and R. Simoniello$^{1,5}$}   

\affil{$^{1}$Instituto de Astrof\'isica de Canarias, C/ V\'ia L\'actea s/n, 38205, La Laguna, 
Tenerife, Spain}   
\affil{$^{2}$Laboratoire AIM, CEA/DSM-CNRS, Universit\'e Paris 7 Diderot, 
IRFU/SAp, Centre de Saclay, 91191, Gif-sur-Yvette, France}
\affil{$^{3}$Indian Institute of Astrophysics, Koramangala, Bangalore 560034, India}
\affil{$^{4}$Observatoire de Bordeaux, B.P. 89, 33270, Floirac cedex, France}
\affil{$^{5}$Physikalisch-Meteorologisches Observatorium Davos/World Radiation 
center, Dorfstrasse 33, 7260 Davos Dorf, Switzerland}

\begin{abstract} 
The primary challenge of GOLF-NG (Global Oscillations at Low Frequency New Generation) is 
the detection of the low-frequency solar gravity and acoustic modes, as well as the possibility to measure the 
high-frequency chromospheric modes. On June 8th 2008, the first sunlight observations 
with the multichannel resonant GOLF-NG prototype spectrometer were obtained at the 
Observatorio del Teide (Tenerife). The instrument performs integrated (Sun-as-a-star), Doppler
velocity measurements, simultaneously at eight different heights in the D1 sodium line profile, 
corresponding to photospheric and chromospheric layers of the solar atmosphere. 
In order to study its performances, to validate the conceived strategy, and to estimate the necessary improvements,
this prototype has been running on a daily basis over the whole summer of 2008 at the Observatorio del 
Teide. We present here the results of the first GOLF-NG observations, clearly showing 
the characteristics of the 5-minute oscillatory signal at different heights in the 
solar atmosphere. We compare these signals with simultaneous observations from GOLF/SOHO and from 
the Mark-I instrument --- a node of the BiSON network, operating at the same site.
\end{abstract}

\section{The multichannel resonant GOLF-NG spectrometer}
The GOLF-NG instrument (Turck-Chi\`eze et al. 2006) is a 15 points resonant scattering spectrophotometer 
observing in the blue and red wings of the D1 sodium line. The sodium cell, where resonant scattering
is produced, is placed in a varying permanent magnet 
(between 0 and 12~kG), and the resonant scattered light (resulting from, alternatively, 
left and right circular polarized incident light) is extracted at eight different positions 
(channels) along the cell, and later collected on a photodiode matrice detector. Four optical fibers are 
placed around the cell for each channel 
position, thus increasing the total photon counting rate compared to other existing 
instruments. The objective is to reduce at maximum all the instrumental noise sources in 
order to increase the sensitivity to solar oscillations. In order 
to validate the instrumental concept and evaluate its current performances, the GOLF-NG prototype 
has been observing the Sun at the Observatorio del Teide (Tenerife) for a 2008 summer campaign. The 
analysis is currently underway and we present here the first preliminary results.   

\begin{figure}
\centering
\plottwo{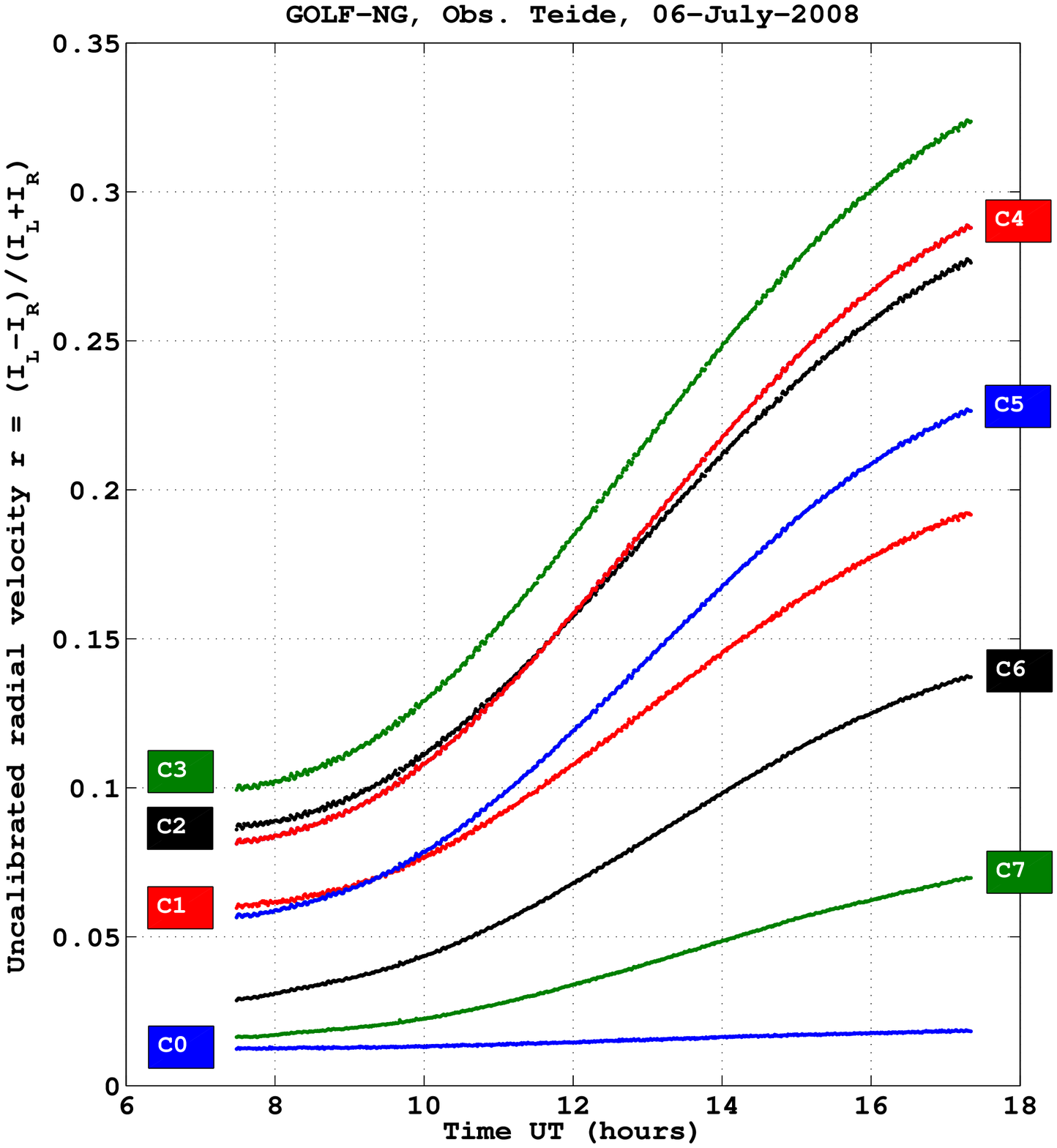}{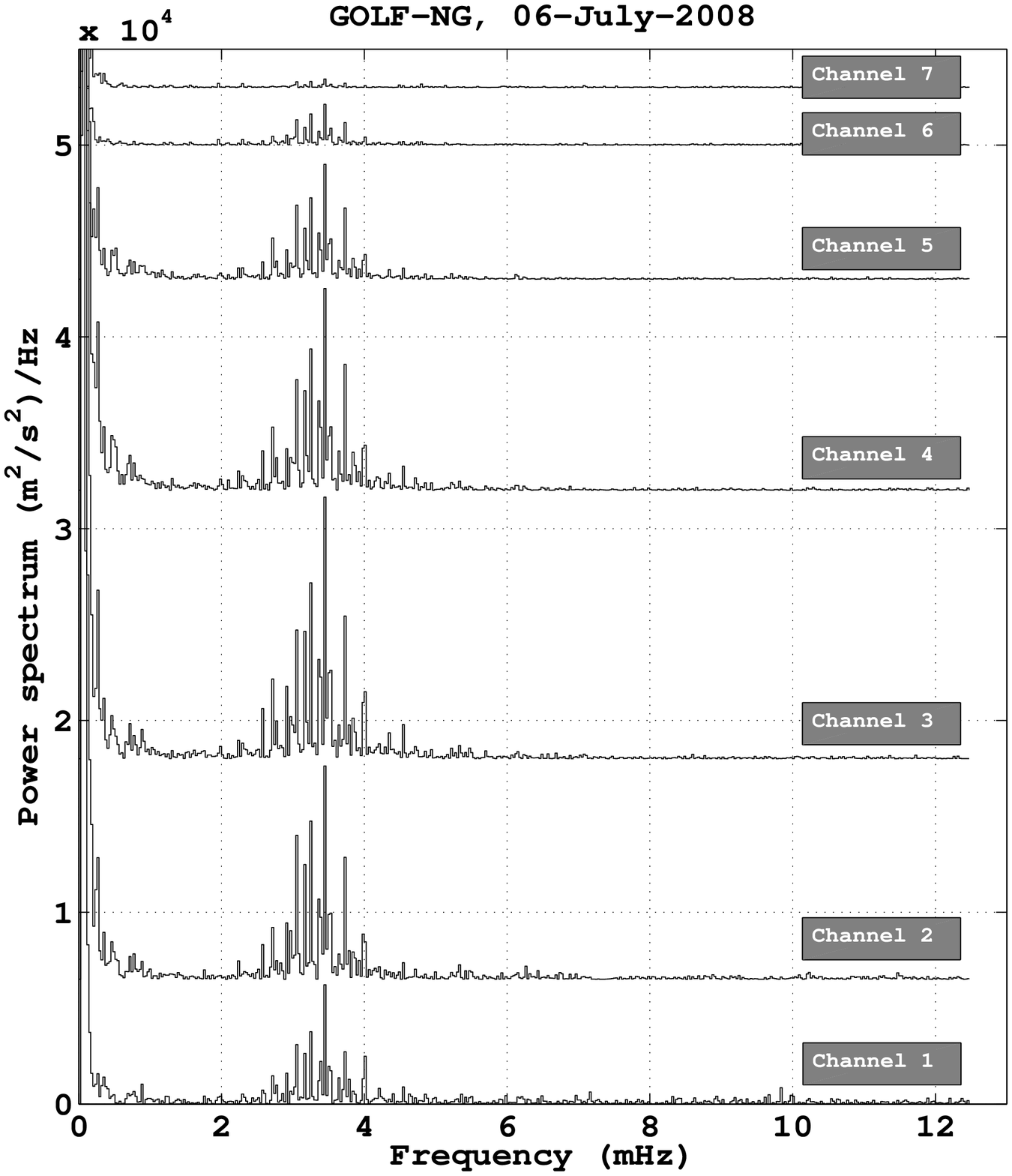}
\caption{Daily uncalibrated radial velocity signal for the different channels of GOLF-NG 
(left panel), and their associated daily power spectra (right panel), for July 06, 2008.}
\label{fig:r_and_pw}
\end{figure}

\section{Solar radial velocity and 5-minute oscillations}
\begin{figure}
\centering
\begin{tabular}{cc}
\resizebox{6.5cm}{!}{\includegraphics[width=1\textwidth]{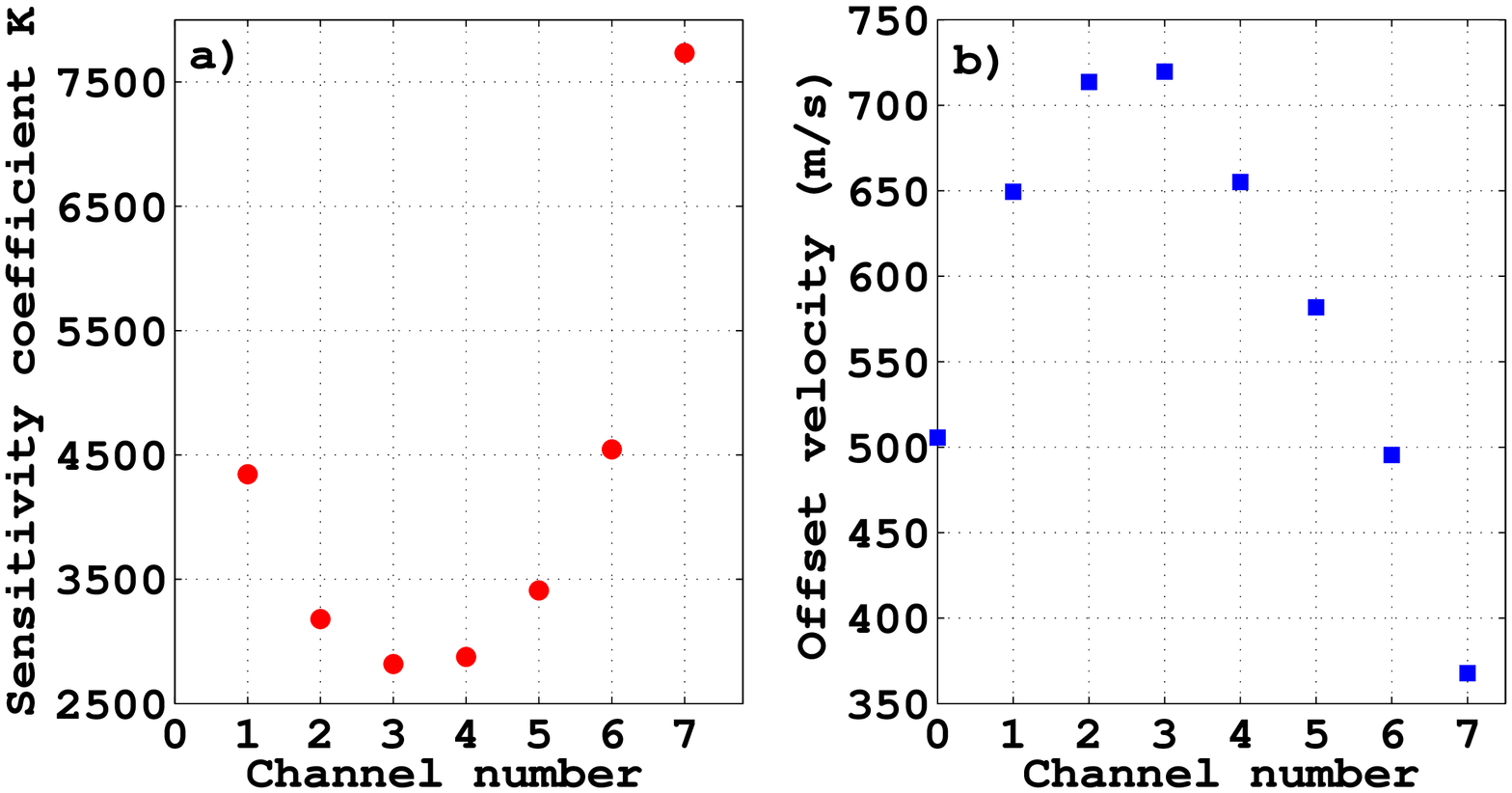}}&    
\resizebox{5.5cm}{!}{\includegraphics[width=1\textwidth]{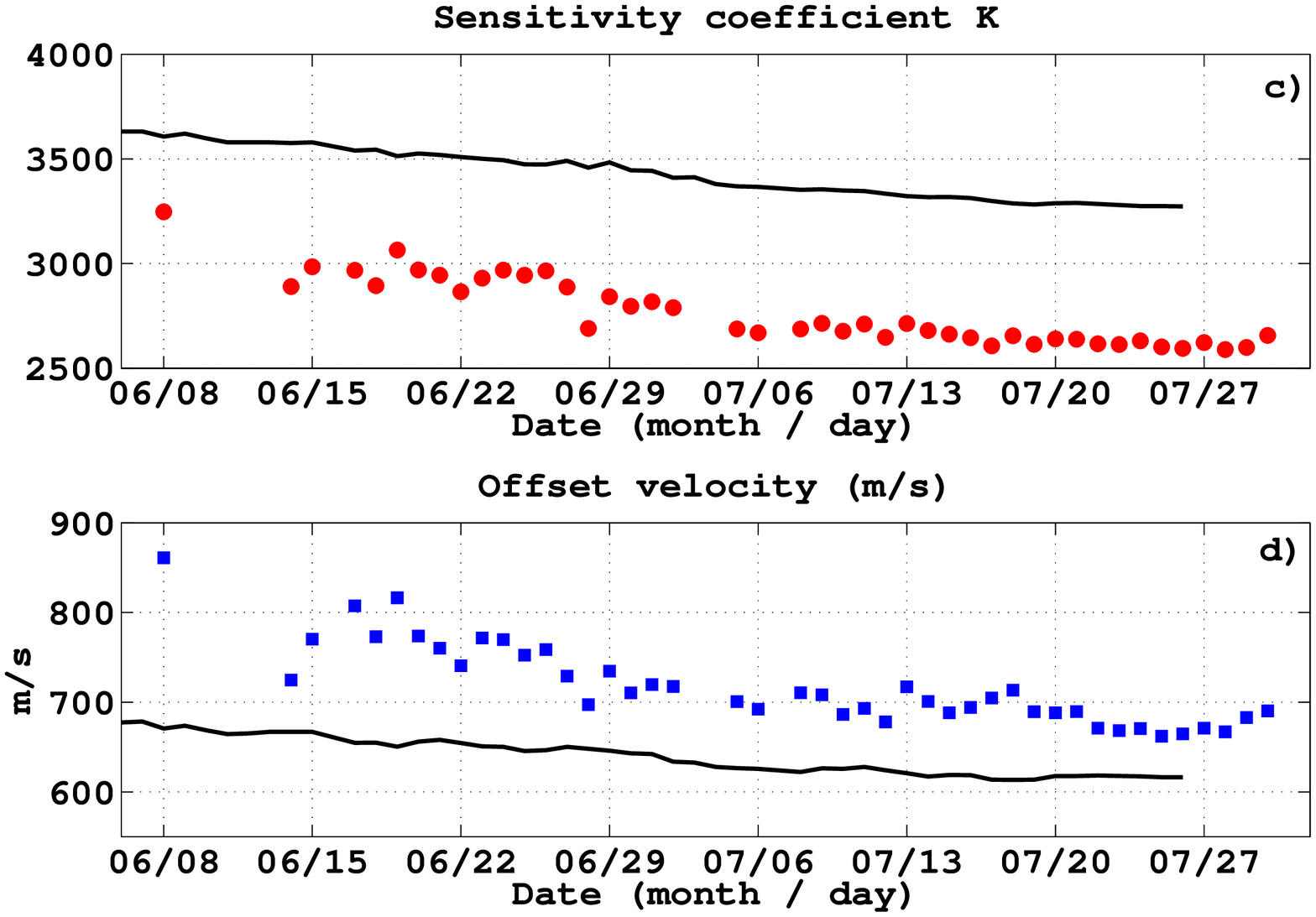}}     
\end{tabular}        
\caption{Sensitivity coefficients K (a) and offset velocities (b) as a function
of the GOLF-NG channels. Daily evolution of the sensitivity coefficients K (c) and offset velocities 
(d) for the GOLF-NG channel 4 along the observing run. The black lines show the corresponding 
values for Mark-I data, applying the same simplified calibration process.}
\label{fig:kcal}
\end{figure}

The uncalibrated solar radial velocity $r = (I_L - I_R) / (I_L + I_R)$ is computed
at the different channel positions of GOLF-NG. $I_L$ and $I_R$ are the number of photons 
collected by the detectors coming (with a 1-second cadence) from the left and right wings 
of the NaD1 line respectively. Figure \ref{fig:r_and_pw} shows the uncalibrated ratio $r$ and
the associated power spectra for the GOLF-NG channels obtained for a daily run of observations.
The observed power of the 5-minute oscillations varies as we move along the NaD1 line, i.e. as we
observe at different heights in the solar atmosphere, although the main pattern (5-minute band) is similar.
The daily residual velocities are obtained by fitting the observations as a 
function of the total line-of-sight velocity computed from the ephemerides. 
The output parameters of this procedure are the coefficient K (instrumental sensitivity) 
and the offset velocity (representative to the instrumental stability), and are represented as a function
of the GOLF-NG channels on  Figs.~\ref{fig:kcal}a and \ref{fig:kcal}b. 
The daily evolution of the GOLF-NG instrument (Figs.~\ref{fig:kcal}c and \ref{fig:kcal}d)
presents a high day-to-day stability in addition to the expected decreasing trend
resulting from the (provisional) lack of proper line profile corrections in the calibration process
(Pall\'e et al. 1993).


\section{Simultaneous observations from GOLF-NG, GOLF, and Mark-I}
\begin{figure}
\centering
\plottwo{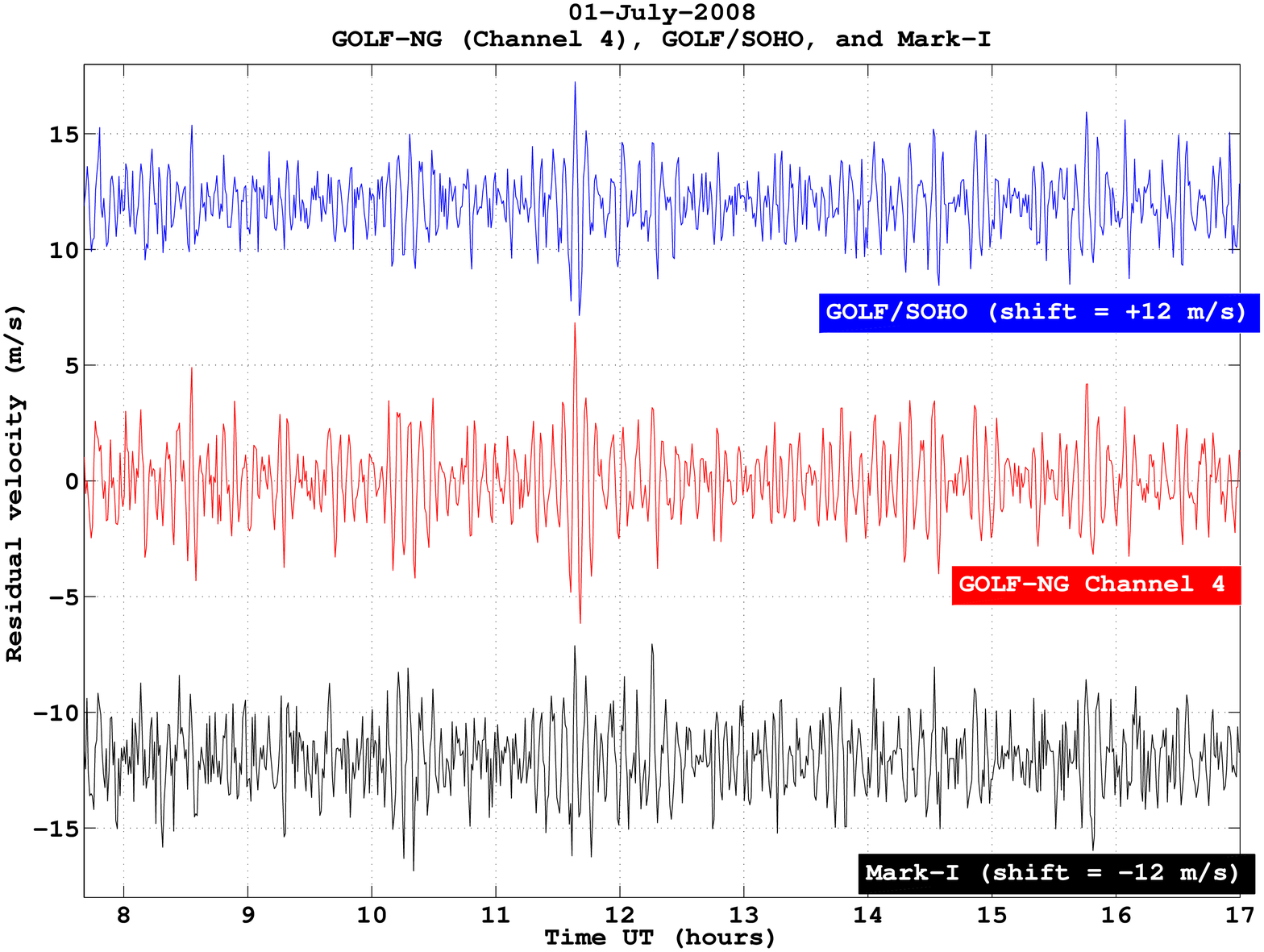}{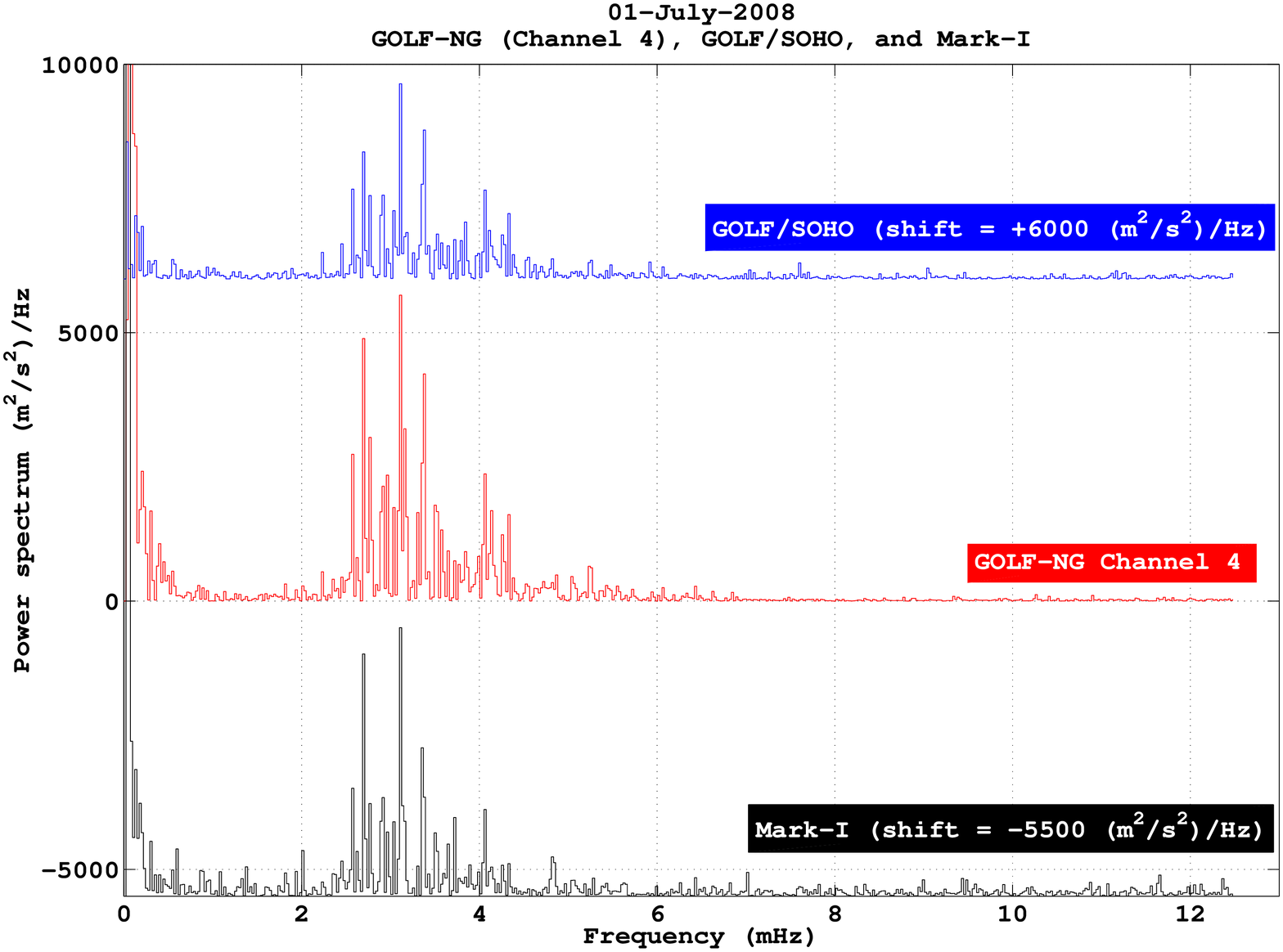}
\caption{Daily residual velocities (m/s) and corresponding power spectra showing the 
5-minute oscillations observed on July 01, 2008 by the instruments GOLF-NG (Obs. del Teide) 
for channel 4 (approximatively same height as GOLF), GOLF on-board SOHO, both observing 
in the Na line, and Mark-I (Obs. del Teide) observing in the K line.}
\label{fig:compa}
\end{figure}

Figure~\ref{fig:compa} compares GOLF-NG observations with simultaneous
observations collected by the space-based GOLF/SOHO instrument observing 
in the Na(D1+D2) lines (approximatively the same height as GOLF-NG channel 4), and by the
Mark-I instrument --- a node of the BiSON network, operating at the same site
as GOLF-NG and observing in the K line.

\section{Photon noise: GOLF-NG (Obs. del Teide) and GOLF/SOHO}
\begin{figure}
\centering
\includegraphics[width=0.7\textwidth]{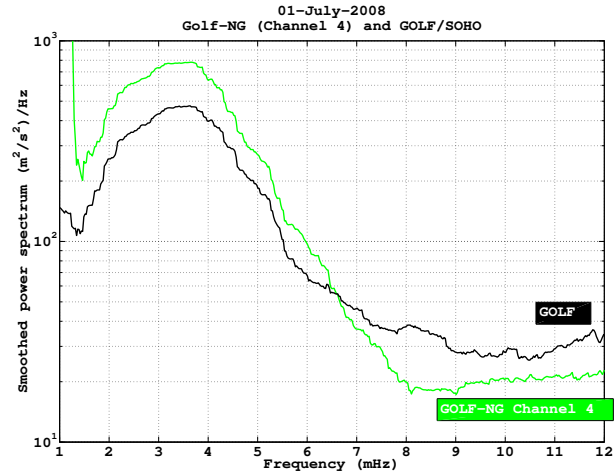}
\caption{Smoothed power spectra from  $\approx$~10 hours of simultaneous observations from GOLF-NG channel~4 (green) and GOLF/SOHO (black).}
\label{fig:photon}
\end{figure}

Figure~\ref{fig:photon} shows simultaneous spectra observed by the ground-based GOLF-NG and space-based 
GOLF/SOHO instruments during $\approx$~10 hours on July 01, 2008. The use of multiple 
fibres at each channel position in GOLF-NG results in a lower high-frequency noise 
(above 8~mHz, region dominated by the photon noise) and in a better signal-to-noise ratio in 
the p-mode range (of a factor $\approx$~3) compared to the present GOLF/SOHO instrument after 12 years
 of operation (Garc\'\i a et al. 2005). 

\section{Summary}
In this preliminary analysis, we showed that the signature of the 5-minute 
solar oscillations is observed in the different channels of the GOLF-NG instrumental 
prototype. The high-frequency noise level is lower than the one from the present 
GOLF/SOHO observations. 
The optimal way to combine the signal collected by the different channels of GOLF-NG, 
in order to reduce the noise level at low frequency (g-mode region), is currently under
investigation, 
as well as the study of the main instrumental noise sources and their respective contributions. 
If not limited by atmospheric conditions due to the ground-based location of the prototype and by other instrumental factors, 
we will use the observations from this 2008 summer campaign to study the contributions of the solar noise as a function of the height in the 
solar atmosphere in the low-frequency range of the power spectra. 
Technical and instrumental improvements (Turck-Chi\`eze et al. 2008) are in progress 
in order to reach the nominal performances necessary for the study of gravity modes 
and the solar atmosphere characteristics.

\acknowledgements 

We thank the technical and instrumental support received from the different IAC's Services,
as well as the Operator Team from the Observatorio del Teide, with a special thanks to 
A.~Pimienta. This work utilizes data from the Mark-I instrument, a node of the BiSON network. 
D.~S. acknowledges the support of the Spanish grant PNAyA2007-62650.

\end{document}